\documentclass[apj]{emulateapj}
\usepackage{natbib}

\def\kms{\,km\,s$^{-1}$}       %km.s -1
\def\vsini{$v$\,sin\,$i$}      %vsini
\def\Msol{$M_\odot$}		%Msun
\def\Rsol{$R_\odot$}		%Rsun
\def\Psol{$\rho_\odot$}		%Psol
\def\Rstar{$R_*$}		%Rstar
\def\Rplanet{$R_{\rm P}$}	%Rplanet
\def\Mjup{$M_{\rm Jup}$}	%Mjup
\def\Rjup{$R_{\rm Jup}$}	%Rjup
\def\teql{$T_{\rm eql}$}
\def\teff{$T_{\rm eff}$}
\def\feh{[Fe/H]}
\def\logg{$\log g_*$}
\def\vsini{$v \sin i$}
\def\mictrb{$\xi_{\rm t}$}
\def\mactrb{$v_{\rm mac}$}

\def\halpha{$H_\alpha$}
\def\ali{$A_{\rm Li}$}		%Lithium abundance
\def\kms{km\, s$^{-1}$}

\def\ecos{$e \cos \omega$}
\def\esin{$e \sin \omega$}
\def\secos{$\sqrt{e} \cos \omega$}
\def\sesin{$\sqrt{e} \sin \omega$}
\def\farcm{\ensuremath{.\mkern-4mu^\prime}}%    % fractional arcminute symbol: 0.'0
\shorttitle{WASP-30b: a 61\,\Mjup\ brown dwarf}
\shortauthors{D. R. Anderson et al.}

\begin{document}

\title{WASP-30\lowercase{b}: a 61\,\Mjup\ brown dwarf transiting a V=12, F8 
star}

\author{D.~R.~Anderson\altaffilmark{1}, 
A.~Collier~Cameron\altaffilmark{2}, 
C.~Hellier\altaffilmark{1}, 
M.~Lendl\altaffilmark{3}, 
P.~F.~L.~Maxted\altaffilmark{1}, 
D.~Pollacco\altaffilmark{4}, 
D.~Queloz\altaffilmark{3}, 
B.~Smalley\altaffilmark{1}, 
A.~M.~S.~Smith\altaffilmark{1}, 
I.~Todd\altaffilmark{4}, 
A.~H.~M.~J.~Triaud\altaffilmark{3}, 
R.~G.~West\altaffilmark{5}, 
S.~C.~C.~Barros\altaffilmark{4}, 
B.~Enoch\altaffilmark{2}, 
M.~Gillon\altaffilmark{6}, 
T.~A.~Lister\altaffilmark{7}, 
F.~Pepe\altaffilmark{3},
D.~S\'egransan\altaffilmark{3},
R.~A.~Street\altaffilmark{7}, 
S.~Udry\altaffilmark{3}}
\email{dra@astro.keele.ac.uk}

\altaffiltext{1}{Astrophysics Group, Keele University, Staffordshire, ST5 5BG, 
UK}
\altaffiltext{2}{SUPA, School of Physics and Astronomy, University of St. Andrews, 
North Haugh, Fife, KY16 9SS, UK}
\altaffiltext{3}{Observatoire de Gen\`eve, Universit\'e de Gen\`eve, 51 Chemin 
des Maillettes, 1290 Sauverny, Switzerland}
\altaffiltext{4}{Astrophysics Research Centre, School of Mathematics \& Physics,
 Queen's University, University Road, Belfast, BT7 1NN, UK}
\altaffiltext{5}{Department of Physics and Astronomy, University of Leicester, 
Leicester, LE1 7RH, UK}
\altaffiltext{6}{Institut d'Astrophysique et de G\'eophysique,  Universit\'e de 
Li\`ege,  All\'ee du 6 Ao\^ut, 17,  Bat.  B5C, Li\`ege 1, Belgium}
\altaffiltext{7}{Las Cumbres Observatory, 6740 Cortona Dr. Suite 102, Santa 
Barbara, CA 93117, USA}

\begin{abstract}
We report the discovery of a 61-Jupiter-mass brown dwarf, which 
transits its F8V host star, WASP-30, every 4.16 days. 
From a range of age indicators we estimate the system age to be 1--2 Gyr. 
We derive a radius ($0.89 \pm 0.02$ \Rjup) for the companion that is consistent 
with that predicted (0.914 \Rjup) by a model of a 1-Gyr-old, non-irradiated 
brown dwarf with a dusty atmosphere.
The location of WASP-30b in the minimum of the mass-radius relation is 
consistent with the quantitative prediction of \citet{2000ARA&A..38..337C}, thus 
confirming the theory. 
\end{abstract}


\section{Introduction}
A brown dwarf (BD) is traditionally defined as an object with a mass above 
the deuterium-burning limit \citep[13 \Mjup; e.g.][]{2000ApJ...542L.119C} and 
below the hydrogen-burning limit 
\citep[0.07 \Msol; e.g.][]{2000ApJ...542..464C}. 
However, an alternative suggestion is that the manner in which an object 
forms should determine whether it is a planet or a BD. 
Thus, if an object formed by core accretion of dust and ices in a 
protoplanetary disc then it would be a planet, and if it formed by 
gravoturbulent collapse of a molecular cloud, as do stars, then it would be a 
BD. 

Studies such as the \citet{2007A&A...470..903C} observations of a young open 
cluster core find a continuous mass 
function down to $\sim$6~\Mjup, indicating that the star 
formation mechanism can produce objects with planetary masses. 
This is supported by theoretical studies 
\citep{2004ApJ...617..559P,2008ApJ...684..395H} which suggest that 
gravoturbulent fragmentation of molecular clouds produces stars and BDs down to 
a few Jupiter masses in numbers comparable to the observationally-determined 
distribution.
In contrast, when taking into account planetary migration through the 
protoplanetary disc, the core accretion process might result in giant planets 
with masses of up to 10~\Mjup\ \citep{2005A&A...434..343A}, or even 25~\Mjup\ 
\citep{2008ASPC..398..235M}. 
\citet{2010arXiv1009.5991S} see evidence for a a bimodal distribution in brown 
dwarf masses, with the less-massive group presumably representing the high-mass 
tail of the planetary distribution. 

An accurate, precise measurement of an object's radius is therefore required to 
probe for the existence of a core and thus discriminate between the two 
formation mechanisms. For example, the radius of the 8-\Mjup\ body, 
HAT-P-2b, is consistent with an irradiated planet incorporating a 
340-Earth-mass core, but is smaller than if it were coreless 
\citep{2009A&A...506..385L}. 
The 22-\Mjup\ CoRoT-3b \citep{2008A&A...491..889D} is sufficiently massive to 
qualify as a BD under the traditional definition, 
but the radius of this object is uncertain at the 7\% level. This is higher than 
the 3\% required to discriminate between the absence or 
presence of a core and thus determine how it formed \citep{2009A&A...506..385L}. 
\citet{2010ApJ...718.1353I} found a $\sim$30-\Mjup\ BD, 
NLTT 41135C, which transits one member of an M-dwarf binary system. 
However, as the transits are grazing, it is not currently possible to accurately 
measure its radius. 

There is less ambiguity around the upper end of the BD mass regime: if 
a body is sufficiently massive to fuse hydrogen then it is a star, otherwise it 
is a BD. High-mass BDs with precise radius measurements are 
useful for testing BD evolution models, as it is in the high-mass 
regime that models predict the greatest changes in radius with age 
\citep[e.g.][]{2003A&A...402..701B}. 
\citet{2006Natur.440..311S} discovered a BD eclipsing binary system in 
the Orion Nebula star-forming region, with masses of $57 \pm 5$ \Mjup\ and 
$36 \pm 3$ \Mjup. With very large radii of $0.699 \pm 0.034$ \Rsol\ and 
$0.511 \pm 0.026$ \Rsol, it seems that these objects are in the earliest stages 
of gravitational contraction. 
Similar to the NLTT 41135 system, LHS~6343~C 
\citep[][Johnson 2010, private communication]{2010arXiv1008.4141J} is a 
63-\Mjup\ BD that transits one member of an M-dwarf binary system. 
In this case, the transits are full and so the radius ($0.825 \pm 0.023$ \Rjup) 
of this object is precisely determined. 
CoRoT-15b \citep{2010arXiv1010.0179B} is a 63-\Mjup-mass 
BD in a 3-d orbit around an F7V star. Due to the faintness of the host 
star ($V \sim 16$), the BD radius ($1.12^{+0.30}_{-0.15}$ \Rjup) is not yet well 
determined.

To test and refine models of BD formation and evolution, a population 
of well-characterised objects is required. 
In this letter, we present the discovery of \objectname{WASP-30b}, a 61-\Mjup\ 
brown dwarf that transits its moderately-bright host star. 

\section{Observations}
\begin{figure}
\centering                     
\includegraphics[width=9cm]{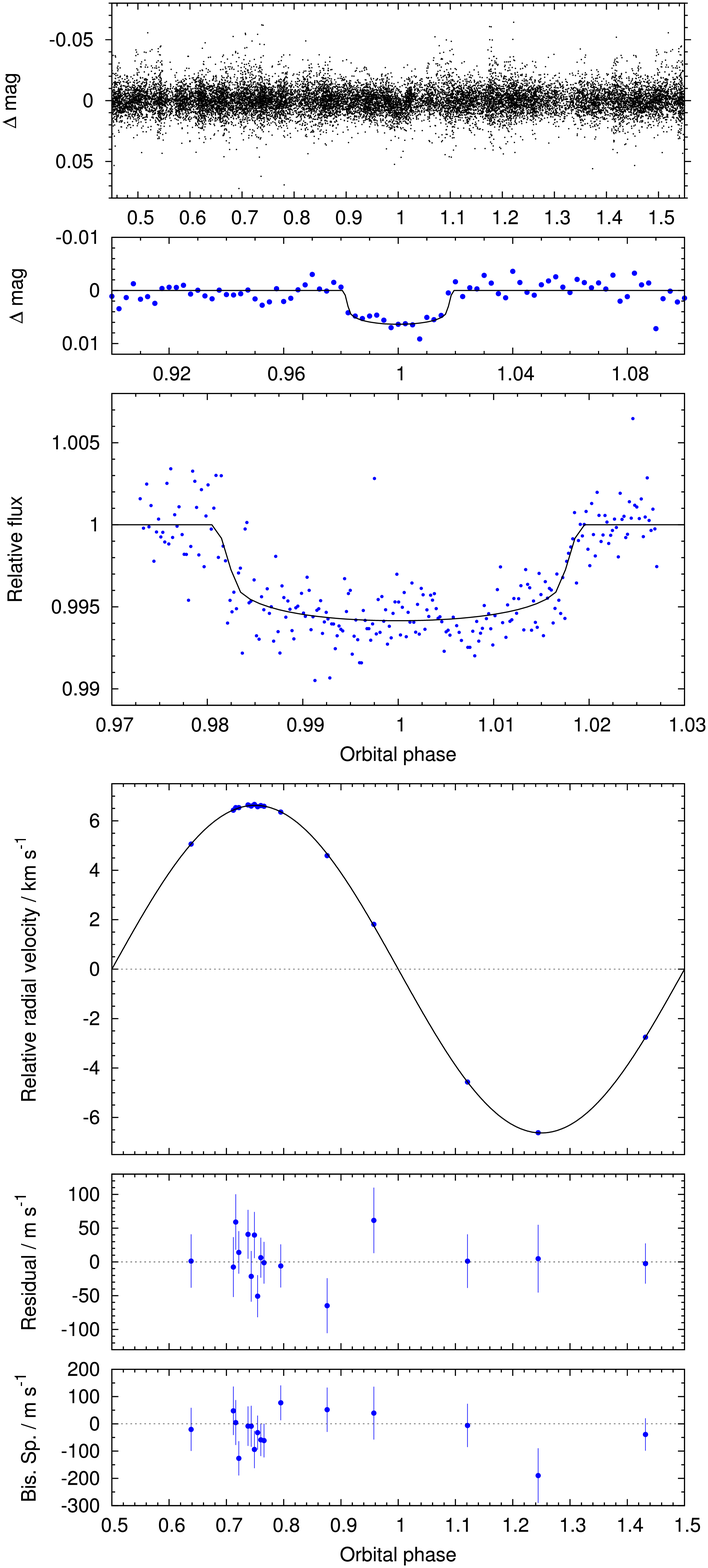}
\caption{  
{\bfseries{\em Top panel}}: WASP discovery light curve, folded on the ephemeris 
of Table~\ref{tab:mcmc}. 
{\bfseries{\em Second panel}}: Binned ($\Delta \phi = 0.0025$) WASP data around 
the transit.
{\bfseries{\em Third panel}}: Transit light curve obtained with Euler. 
{\bfseries{\em Fourth panel}}: relative RV measurements of WASP-30 
measured by CORALIE. 
{\bfseries {\em Fifth panel}}: residuals about the model RV solution.
{\bfseries {\em Bottom panel}}: bisector spans.
\label{fig:phot-rv}}
\end{figure}

\objectname{WASP-30} is a $V = 11.9$, F8V star located in Aquarius, on the 
border with Cetus. 
A transit search \citep{2006MNRAS.373..799C} of WASP-South data from 2008 July 
to November found a strong, 4.16-d periodicity.  Further observations in 2009 
with both WASP instruments \citep{2006PASP..118.1407P} led to a total of 17\,612 
usable photometric measurements (Figure~\ref{fig:phot-rv}).

Using the CORALIE spectrograph mounted on the 1.2-m Euler-Swiss telescope 
\citep{1996A&AS..119..373B,2000A&A...354...99Q}, we obtained 16 spectra of 
WASP-30 in 2009.
Radial velocity (RV) measurements were computed by weighted cross-correlation 
\citep{1996A&AS..119..373B,2005Msngr.120...22P} with a numerical G2-spectral 
template.
RV variations were detected with the same period found from the WASP photometry 
and with a semi-amplitude of 6.6 km s$^{-1}$, consistent with a sub-stellar-mass 
companion. The RV measurements are listed in Table~\ref{tab:rv} and are plotted 
in Figure~\ref{fig:phot-rv}.

%%%%%%%%%%%%%%%%%%%%%% 
%% RV data table 
%%%%%%%%%%%%%%%%%%%%%% 
\begin{table} 
\centering
\caption{Radial velocity measurements of WASP-30\label{tab:rv}} 
\begin{tabular*}{0.4\textwidth}{@{\extracolsep{\fill}}cccc} 
\hline 
BJD--2\,450\,000 & RV & $\sigma$$_{\rm RV}$ & BS\\ 
(days) & (km s$^{-1}$) & (km s$^{-1}$) & (km s$^{-1}$)\\ 
\hline
5009.9065 & 14.275 & 0.032 & \,\,~0.077\\
5040.8722 & \,~1.298 & 0.050 & $-$0.190\\
5092.6977 & 14.348 & 0.044 & \,\,~0.048\\
5095.6894 & \,~5.163 & 0.030 & $-$0.039\\
5096.5476 & 12.979 & 0.040 & $-$0.020\\
5096.8712 & 14.452 & 0.041 & \,\,~0.005\\
5097.5351 & 12.509 & 0.041 & \,\,~0.052\\
5097.8735 & \,~9.732 & 0.048 & \,\,~0.039\\
5098.5538 & \,~3.350 & 0.040 & $-$0.006\\
5113.5209 & 14.451 & 0.031 & $-$0.127\\
5113.5877 & 14.563 & 0.036 & $-$0.008\\
5113.6111 & 14.515 & 0.038 & $-$0.009\\
5113.6343 & 14.582 & 0.034 & $-$0.094\\
5113.6576 & 14.488 & 0.031 & $-$0.032\\
5113.6809 & 14.535 & 0.030 & $-$0.059\\
5113.7041 & 14.508 & 0.031 & $-$0.061\\
\hline 
\end{tabular*} 
\end{table}

To test the hypothesis that the RV variations are due to spectral line 
distortions caused by a blended eclipsing binary or starspots, we performed a 
line-bisector analysis \citep{2001A&A...379..279Q} of the CORALIE 
cross-correlation functions. 
The lack of correlation between bisector span and RV 
(Figure~\ref{fig:phot-rv}) supports our conclusion that the periodic dimming of 
WASP-30 and its RV variations are due to the sub-stellar orbiting body, 
WASP-30b.

To refine the system parameters, we obtained high signal-to-noise transit 
photometry through a Gunn $r$ filter with the Euler-Swiss telescope on 
2010 Aug 01 (Figure~\ref{fig:phot-rv}; Table~\ref{tab:euler}). 
The data were affected by light cloud and a guiding issue caused by the close 
proximity of the bright Moon (69\% illumination, 26$^\circ$ from target). 

%%%%%%%%%%%%%%%%%%%%%% 
%% Euler photometry table 
%%%%%%%%%%%%%%%%%%%%%% 
\begin{table} 
\centering
\caption{Euler photometry of WASP-30\label{tab:euler}} 
\begin{tabular*}{0.4\textwidth}{@{\extracolsep{\fill}}ccc} 
\hline 
BJD--2\,450\,000 & Relative flux & $\sigma$$_{\rm flux}$\\ 
(days) & & \\ 
\hline
5409.693659	& 1.00123	& 0.00185	\\
5409.695116	& 0.99944	& 0.00185	\\
5409.696482	& 1.00213	& 0.00185	\\
\ldots		& \ldots	& \ldots	\\
5409.917840	& 1.00020	& 0.00185	\\
5409.918650	& 0.99791	& 0.00185	\\
\hline 
\end{tabular*}
\\This table is published in its entirety in the electronic edition of the 
{\it Astrophysical Journal}.  A portion is shown here for guidance regarding 
its form and content.
\end{table}

\section{Stellar parameters}
The 16 CORALIE spectra of WASP-30 were co-added to produce a
single spectrum with a typical S/N of around 70:1. 
The analysis was performed using the methods given in 
\citet{2009A&A...501..785G}.
The \halpha\ line was used to determine the
effective temperature (\teff), while the Na {\sc i} D and Mg {\sc i} b lines
were used as surface gravity (\logg) diagnostics. The parameters obtained from
the analysis are given in the top panel of Table~\ref{tab:mcmc}. 
The elemental abundances were determined from equivalent width measurements of 
several clean and unblended lines. A value for microturbulence (\mictrb) was 
determined from Fe~{\sc i} using the method of \citet{1984A&A...134..189M}. The 
quoted error estimates include that given by the uncertainties in \teff, \logg\ 
and \mictrb, as well as the scatter due to measurement and atomic data 
uncertainties. 
Our quoted lithium abundance takes account of non local thermodynamic 
equilibrium corrections \citep{1994A&A...288..860C}, with a value of 
\ali\ = 2.95 resulting when neglecting them. 

The projected stellar rotation velocity (\vsini) was determined by fitting the
profiles of several unblended Fe~{\sc i} lines. A value for macroturbulence
(\mactrb) of 4.7 $\pm$ 0.3 \kms\ was assumed, based on the tabulation by 
\citet{2008oasp.book.....G}, and an instrumental FWHM of 0.11 $\pm$ 0.01 \AA, 
determined from the telluric lines around 6300\AA. A best-fitting value of 
\vsini\ = 14.2 $\pm$ 1.1~\kms\ was obtained.

\begin{figure}
\centering                     
\includegraphics[width=9cm]{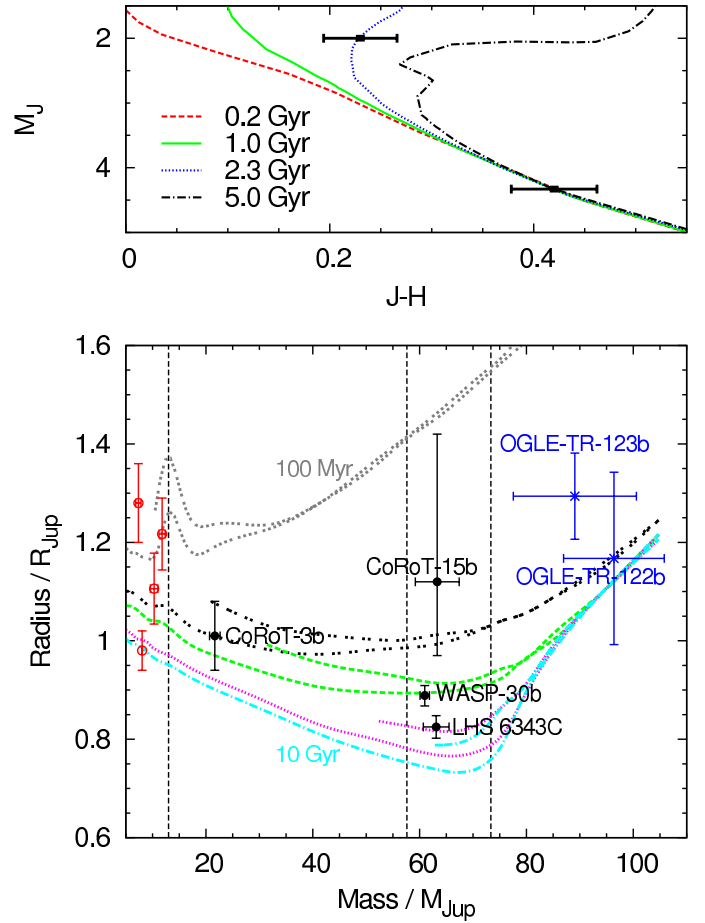}
\caption{
{\bfseries{\em Top panel}}: 
Colour-magnitude diagram for WASP-30 and a 
nearby star with which it appears to be comoving. Isochrones 
for the ages shown are from \citet{2008A&A...482..883M}. 
{\bfseries{\em Bottom panel}}: 
Mass-radius diagram showing objects in the mass range 5--110~\Mjup\ with precise 
radius measurements. The data are from http://www.inscience.ch/transits/ and, 
in the case of LHS~6343~C, from Johnson (2010, private communication). 
The theoretical isochrones with ages 0.1, 0.5, 1, 5 and 10 Gyr are from the 
DUSTY00 and COND03 models. Those that reach down to the minimum 
plotted mass are the COND03 isochrones. 
The vertical dashed lines depict the approximate theoretical minimum masses for 
deuterium burning (13 \Mjup), 
lithium burning (0.055 \Msol), 
and hydrogren burning \citep[0.07 \Msol; e.g.][]{2000ApJ...542L.119C, 
2000ApJ...542..464C}.
\label{fig:colmag-mr}}
\end{figure}

\section{System parameters}
\label{sys-par}
The WASP and Euler photometry were combined with the CORALIE RV measurements in 
a simultaneous Markov-chain Monte Carlo (MCMC) analysis 
\citep{2007MNRAS.380.1230C, 2008MNRAS.385.1576P}. 
Our proposal parameters are: $T_{\rm c}$, $P$, $\Delta F$, $T_{14}$, 
$b$, $K_{\rm 1}$, \teff, \feh, \secos\ and \sesin\ 
\citep{2007MNRAS.380.1230C, 2010A&A...516A..33E}. 
Here $T_{\rm c}$ is the epoch of mid-transit, $P$ is the orbital period, 
$\Delta F = R_{\rm p}^2/R_*^2$ is the fractional flux-deficit that would be 
observed during transit 
in the absence of limb-darkening, $T_{14}$ is the total transit duration (from 
first to fourth contact), $b$ is the impact parameter of the BD's path 
across the stellar disc, $K_{\rm 1}$ is the semi-amplitude of the stellar 
reflex velocity, \teff\ is the stellar effective temperature, \feh\ is the 
stellar metallicity, $e$ is the orbital eccentricity and $\omega$ is the 
argument of periastron. 

As \citet{2006ApJ...642..505F} notes, it is convenient to use 
$e\cos\omega$ and $e\sin\omega$ as MCMC jump parameters, because these two 
quantities are nearly orthogonal and their joint probability density function is 
well-behaved when the eccentricity is small and $\omega$ is highly uncertain. 
Ford cautions, however, that the use of \ecos\ and \esin\ as jump parameters
implicitly imposes a prior on the eccentricity that increases linearly with $e$.
Instead we use \secos\ and \sesin\ as jump parameters,
which restores a uniform prior on $e$.

At each step in the MCMC procedure, each proposal parameter is perturbed from
its previous value by a small, random amount. 
From the proposal parameters, model light and RV curves are generated and 
$\chi^{2}$ is calculated from their comparison with the data. 
A step is accepted if $\chi^{2}$ (our merit function) is lower than for the 
previous step, and a step with higher $\chi^{2}$ is accepted with probability 
$\exp(-\Delta \chi^{2}/2)$. 
In this way, the parameter space around the optimum solution is thoroughly 
explored.
The value and uncertainty for each parameter are respectively taken as the 
median and central 68.3\% confidence interval of the parameter's marginalised 
posterior probability distribution.

From the proposal parameters, we calculate the mass $M$, radius $R$, density 
$\rho$, and surface gravity $\log g$ of the star (which we denote with subscript 
*) and the planet (which we denote with subscript P). 
At each step, the stellar density is measured from the transit light curve 
\citep{2003ApJ...585.1038S}. This is input in to the empirical mass calibration 
of \cite{2010A&ARv..18...67T}, as modified by \citet{2010A&A...516A..33E}, 
to obtain an estimate of the stellar mass. 
We also calculate the equilibrium temperature of the planet \teql, assuming it 
to be a black-body with efficient redistribution of energy from the planet's 
day-side to its night-side, the transit ingress and egress durations, 
$T_{\rm 12}$ and $T_{\rm 34}$, and the orbital semi-major axis $a$. 

With eccentricity floating, we find $e = 0.0021^{+0.0024}_{-0.0015}$.
Applying the `F-test' of \citet{1971AJ.....76..544L}, we find a 70\% 
probability that the fitted eccentricity could have arisen by chance if the 
the underlying orbit is in fact circular. 
As such, we impose a circular orbit, but we note that doing so 
has no significant effect on the solution in this case.

Without exquisite photometry, our implentation of MCMC tends to bias 
the impact parameter, and thus \Rstar\ and \Rplanet, to higher values. 
This is because, with low signal-to-noise photometry, the transit ingress and 
egress durations are uncertain, and symmetric uncertainties in those translate 
into asymmetric uncertainties in $b$ and thus on \Rstar. 
We therefore place a main sequence (MS) prior on the star, which is 
reasonable given the star's apparent age (Section~\ref{sys-age}). 
With the MS prior, a Bayesian penalty ensures that, in accepted MCMC steps, the 
values of stellar radius are consistent with the values of stellar mass for a 
main-sequence star \citep{2007MNRAS.380.1230C}.

The median parameter values and their 1-$\sigma$ uncertainties from our MCMC 
analysis are presented in the middle panel of Table~\ref{tab:mcmc}. 
The corresponding transit light curves and RV curve are shown in 
Figure~\ref{fig:phot-rv}. 
When not imposing a MS prior, the best-fitting values we obtain are: 
$b=0.24^{+0.24}_{-0.16}$, \Rstar\ = $1.337^{+0.147}_{-0.042}$ \Rsol, and 
\Rplanet\ = $0.925^{+0.118}_{-0.040}$ \Rjup.

%%%%%%%%%%%%%%%%%%%%%%%
%MCMC paramters 
%%%%%%%%%%%%%%%%%%%%%%%
\begin{table}
\caption{System parameters for WASP-30} 
\label{tab:mcmc} 
\begin{tabular*}{0.48\textwidth}{@{\extracolsep{\fill}}lc} 
\hline
\multicolumn{2}{l}{Stellar parameters from spectroscopic analysis}\\
\hline
\multicolumn{2}{l}{R.A. = 23$\rm^{h}$53$\rm^{m}$38.03$\rm^{s}$, Dec. = --10$^\circ$07$^{'}$05.1$^{''}$ (J2000)}\\
\multicolumn{2}{l}{TYC 5834-95-1, 2MASS 23533805-1007049}\\
\teff\ (K)	& 6100 $\pm$ 100	\\
\logg		& 4.3 $\pm$ 0.1		\\
\mictrb\ (\kms)	& 1.1 $\pm$ 0.1		\\
\vsini\ (\kms)	& 14.2 $\pm$ 1.1	\\
{[Fe/H]}   	&$-$0.08 $\pm$ 0.10 	\\
{[Si/H]}   	& +0.04 $\pm$ 0.13 	\\
{[Ca/H]}   	& +0.10 $\pm$ 0.14 	\\
{[Ti/H]}   	&$-$0.01 $\pm$ 0.14 	\\
{[Ni/H]}   	&$-$0.08 $\pm$ 0.13 	\\
\ali		& 2.87 $\pm$ 0.10	\\
\hline
\multicolumn{2}{l}{Parameters from MCMC analysis}\\
\hline 
$P$ (d) & $4.156736 \pm 0.000013$\\
$T_{\rm c}$ (HJD) & $2455334.98479 \pm 0.00076$\\
$T_{\rm 14}$ (d) & $0.1595 \pm 0.0017$\\
$T_{\rm 12}=T_{\rm 34}$ (d) & $0.01060 \pm 0.00025$\\
$\Delta F=R_{\rm P}^{2}$/R$_{*}^{2}$ & $0.00498 \pm 0.00017$\\
$b$ \smallskip & 0.066$^{+ 0.072}_{- 0.044}$\\
$i$ ($^\circ$) \medskip & 89.57$^{+ 0.28}_{- 0.47}$\\
$K_{\rm 1}$ (km s$^{-1}$) & $6.627 \pm 0.015$\\
$a$ (AU)  & $0.05325 \pm 0.00039$\\
$e$ & 0 (adopted)\\
$\gamma$ (km s$^{-1}$) \medskip & $7.9177 \pm 0.0099$\\
$M_{\rm *}$ (\Msol) & $1.166 \pm 0.026$\\
$R_{\rm *}$ (\Rsol) & $1.295 \pm 0.019$\\
$\log g_{*}$ (cgs) & $4.280 \pm 0.010$\\
$\rho_{\rm *}$ (\Psol)  & $0.537 \pm 0.019$\\
\teff\  (K) & $6201 \pm 97$\\
\feh \medskip & $-0.03 \pm 0.10$\\
$M_{\rm P}$ ($M_{\rm Jup}$) & $60.96 \pm 0.89$\\
$R_{\rm P}$ ($R_{\rm Jup}$) & $0.889 \pm 0.021$\\
$\log g_{\rm P}$ (cgs) & $5.247 \pm 0.019$\\
$\rho_{\rm P}$ ($\rho_{\rm J}$) & $86.8 \pm 5.7$\\
\teql\ (K) & $1474 \pm 25$\\
\hline 
\end{tabular*}\\
\begin{tabular*}{0.48\textwidth}{@{\extracolsep{\fill}}lcccc}
\multicolumn{5}{l}{Proper motions of WASP-30 and its neighbour}\\
\hline
Star & \multicolumn{2}{c}{UCAC3} & \multicolumn{2}{c}{PPXML}\\
 & $\mu_{\alpha}$ (mas/yr) & $\mu_{\delta}$ (mas/yr) & $\mu_{\alpha}$ (mas/yr) & $\mu_{\delta}$ (mas/yr)\\ 
\hline
WASP-30		& $-20.4 \pm 2.2$ & $-9.0 \pm 2.4$ & $-22.9 \pm 2.2$ & $-9.0 \pm 2.4$\\
neighbour	& $-23.0 \pm 7.2$ & $-8.1 \pm 4.4$ & $-18.9 \pm 3.9$ & $-9.5 \pm 3.9$\\
\hline
\\
\end{tabular*}
\end{table} 

\section{System age and companion radius}
\label{sys-age}
The high lithium abundance (\ali\ = $2.87 \pm 0.10$) found in WASP-30 implies an 
age most likely between that of open clusters such as $\alpha$ Per (50 Myr; 
\ali\ = $2.97 \pm 0.13$) and the Hyades (600 Myr; \ali\ = $2.77 \pm 0.21$), and 
almost certainly younger than 2-Gyr-old open clusters such as NGC 752 
\citep[\ali\ = $2.65 \pm 0.13$; ][]{2005A&A...442..615S}. 

Assuming aligned stellar-spin and planetary-orbit axes, the measured 
\vsini\ of WASP-30 and its derived stellar radius indicate a rotational period 
of $P_{\rm rot} = 4.6 \pm 0.4$~days. 
After removing the transits from the WASP light curves, we searched them for 
evidence of rotational modulation. Though there are periodogram peaks at periods 
of 4.1 d, 4.3 d and 4.7 d, the signal amplitudes are small. 
We thus conclude that there is no evidence of rotational modulation in the 
WASP-30 light curves, commensurate with expectations based on the star's 
spectral type.
The 4.6 d stellar rotation period is very close to the companion's orbital 
period, suggesting that the two may be synchronized, thus preventing a 
gyrochronological age determination \citep{2007ApJ...669.1167B}. The 
synchronization timescale \citep{1977A&A....57..383Z} for the star is 
$0.23 \pm 0.02$ Gyr. 

We searched within 15\arcmin\ of the sky position of WASP-30 for 
common proper motion stars. The $V = 13.6$ star USNO-B1.0 0800-0674908 is 
13\farcm09 away and
appears to be comoving with WASP-30 (Table~\ref{tab:mcmc}, bottom panel). 
Using 2MASS photometry, we constructed a colour-magnitude diagram. 
A distance modulus of $8.50 \pm 0.05$ ($500 \pm 10$ pc) was required to place 
the comoving star on the main-sequence and suggests that WASP-30 is 
$\sim$2.3-Gyr old (Figure~\ref{fig:colmag-mr}, upper panel).
However, the apparent magnitude and spectral type of WASP-30 suggest a smaller 
distance modulus of $7.9 \pm 0.2$ ($366 \pm 77$ pc). 
As the comoving star may be a mere line-of-sight neighbour, this age 
determination should be treated with caution.

In the lower panel of Figure~\ref{fig:colmag-mr}, WASP-30b is plotted in a 
mass-radius diagram 
together with isochrones of isolated BDs from models with dusty atmospheres 
\citep[DUSTY00,][]{2000ApJ...542..464C} and models with dust-free atmospheres 
\citep[COND03,][]{2003A&A...402..701B}. 
By an age of 1~Gyr, an isolated BD with the mass of 
WASP-30b is expected to have cooled to \teql\ $\sim$ 1700 K, and the transition 
from dusty L-dwarfs to dust-free T-dwarfs is expected to take place at \teql\ = 
1300--1700 K \citep{2003A&A...402..701B}.
Considering this and the fact that WASP-30b is highly irradiated, it is likely 
that the dusty atmosphere models of \citet{2000ApJ...542..464C} are more 
representative. 
Depending on the transition temperature, WASP-30b may remain a dusty L-dwarf for 
the lifetime of its host star, or it may at some point transition to a 
dust-free T-dwarf.

WASP-30b has a mass of $0.05819 \pm 0.00084$ \Msol\ and a radius of 
$0.0914 \pm 0.0022$ \Rsol.
The DUSTY00 models predict the radius of an isolated, 0.06-\Msol\ BD to 
be 0.149, 0.104, 0.094 and 0.084 \Rsol\ at ages of, respectively, 0.1, 0.5, 1 
and 5 Gyr. 
From a simple linear interpolation of the DUSTY00 model values, the measured 
radius of WASP-30b and its 1-$\sigma$ uncertainty suggests its age is 
$2 \pm 1$ Gyr.
The measured radius is inconsistent with the DUSTY00 model for a 0.5-Gyr 
BD at the 5.7-$\sigma$ level and inconsistent with a 5-Gyr BD at the 
3.4-$\sigma$ level, but consistent for a 1-Gyr BD at the 1.2-$\sigma$ level. 
This agreement would be slightly better if the DUSTY00 models took account of 
irradiation. 
\citet{2003A&A...402..701B} find that the irradiation of a dust-free atmosphere, 
at the level of irradiation experienced by WASP-30b, results in radii larger by 
10\% for a 1-\Mjup\ planet, and larger by 7\% for a 10-\Mjup\ planet. 
However, they note that the evolution of dust-free 
atmospheres are more affected by irradiation than are dusty atmospheres, and 
WASP-30b is considerably more massive. 

WASP-30's lithium abundance favours a young system age of 50--600 Myr, though 
lithium is a poor age indicator for an F8 star, and an age 
of up to 2 Gyr is not ruled out.
The apparent rotational synchronisation of the host star places a lower limit of 
200 Myr (though the system may have been born synchronised) and a possible 
companion star suggests an older age of $~$2.3 Gyr.  
Given the BD's measured radius, the DUSTY00 BD model indicates an age of 
$2 \pm 1$ Gyr. 
Taken together, we suggest that an age of 1--2 Gyr is most likely. 

\section{Discussion}
The discovery of WASP-30b heralds the first unambiguous observational 
determination of the mass-radius relation (MRR) in the BD regime, 
and so we have added BDs to white dwarfs and neutron stars in the 
list of quantum-dominated objects with radius determinations. 

\citet{2000ARA&A..38..337C} perfomed the first quantitative theoretical 
calculation of the MRR in the substellar and low-mass star domain, predicting a 
minimum in the MRR at high BD masses (see Section 3.1 of that paper). 
The location of WASP-30b in the MRR minimum is consistent with the quantitative 
prediction of \citet{2000ARA&A..38..337C}, thus confirming the theory. 

Thus far, we know of two other high-mass brown dwarfs that transit stars: 
CoRoT-15b \citep[63 \Mjup;][]{2010arXiv1010.0179B} and 
LHS~6343~C 
\citep[63 \Mjup;][Johnson 2010, private communication ]{2010arXiv1008.4141J}. 
The radius of CoRoT-15b ($1.12^{+0.30}_{-0.15}$ \Rjup) is uncertain and the age 
of the system is currently unconstrained. The faintness of the host star 
($V \sim 16$) makes improving this situation difficult.
LHS~6343~C was found to transit one member of an M-dwarf binary system using 
Kepler photometry \citep[KIC 10002261; e.g.][]{2010Sci...327..977B}.
It initially seemed that LHS~6343~C was larger than predicted for a BD of its 
mass and age \citep{2010arXiv1008.4141J}. However, after a 
re-evaluation of the treatment of the third light in the system, it seems to be 
consistent (Johnson 2010, private communication). 

\citet{2000ApJ...542..464C} predict that BDs with $M < 0.05$ 
\Msol\ do not burn lithium, those with $M > 0.06$ \Msol\ burn essentially all 
their lithium by an age of 0.5 Gyr, and those with an intermediate mass 
($M = 0.055$ \Msol) burn half their lithium by 0.5 Gyr, two-thirds by 1 Gyr, and 
three-quarters by 5 Gyr.
With a mass of $0.0582 \pm 0.0008$ and an age of 1--2 Gyr, WASP-30b is 
likely to have burned most of its supply of lithium. 

WASP-30b has the second smallest companion-to-star size ratio 
($\Delta F = R_{\rm p}^2/R_*^2 = 0.0050$) 
of all sub-stellar bodies so far discovered by ground-based transit surveys. 
The star in the system with the smaller size ratio, HAT-P-11 
\citep[$V$ = 9.6; $\Delta F = 0.0033$;][]{2010ApJ...710.1724B}, is 8 times 
brighter than WASP-30.
As it is far easier to find such an object around a smaller, cooler star, the 
discovery of WASP-30b suggests that high-mass, sub-stellar objects in short 
orbits around cooler stars are rare. 

\acknowledgments
WASP-South is hosted by the South African Astronomical Observatory and  
SuperWASP-N is hosted by the Issac Newton Group on La Palma. We are 
grateful for their ongoing support and assistance. Funding for WASP comes from 
consortium universities and from the UK's Science and Technology Facilities 
Council.
M. Gillon acknowledges support from the Belgian Science Policy Office in the 
form of a Return Grant.

{\it Facilities:} WASP, \facility{Euler1.2m}

\end{document}